\documentstyle[12pt,iopfts]{ioplppt}
\def\PRA{{\it Phys. Rev. A}}
\begin{document}
\title{Coherent states for the hydrogen atom:
discrete and   continuous spectra }
[Coherent states for the hydrogen atom]
\author{S A Pol'shin\ftnote{1}{E-mail: itl593@online.kharkov.ua}}
\date{}
\address{Institute for Theoretical Physics \\
 NSC Kharkov Institute of Physics and Technology \\
 Akademicheskaia St. 1, 61108 Kharkov, Ukraine  }
\jl{1}


\begin{abstract}
We construct the systems of generalised coherent
states for the discrete and continuous spectra of the hydrogen
atom. These systems are expressed in elementary functions and are invariant under the
$SO(3, 2)$ (discrete spectrum) and $SO(4, 1)$ (continuous
spectrum) subgroups of the dynamical symmetry group $SO(4, 2)$
of the hydrogen atom. Both systems of coherent
states are particular cases of the kernel of integral operator
which interwines irreducible representations of the $SO(4, 2)$ group.
\end{abstract}

\section{Introduction}

The problem of constructing the generalised coherent states (CS) for
the hydrogen atom (HA) was first formulated by Schrodinger in 1926
simultaneously with the construction of CS for the harmonic oscillator (HO)
(see historical remarks in~\cite{Cerjan}).
Since then, there have been attempts to solve this problem on the basis
of the Kustaanheimo--Stiefel transformation connecting 4D HO and
3D HA~\cite{Gerry,BDRG}. Within this approach the CS for 4D HO are
constructed,
whereas the CS for 3D HA are obtained from the above CS either using
constraints imposed on a set of parameters~\cite{Gerry} or by integration
over the additional variable~\cite{BDRG}. However, the packets obtained
in~\cite{BDRG} spread with time; even within the time when the packets preserve
their shape, they can be considered as simulating the Kepler motion
for large quantum numbers only and, as it was shown in~\cite{Nandi},
when restricted to a plane. The same statements are also valid for the
states constructed in~\cite{Gerry}.

It is also possible (as it was suggested in~\cite{MalkinBook})
to start from the existence of integrals of motion. Due to the
dynamical symmetry group $SO(4, 2)$ of the HA~\cite{MM}, this idea naturally leads to
the construction of CS by the Barut--Girardello method, as it was
performed in~\cite{Bracken}. Such states evolve consistently under the
pseudo-Hamiltonian $(-2H)^{-1/2}$; however, during evolution in real time
they spread and cannot be expressed in a closed form in the configuration
space. Much progress has been achieved in constructing the CS for the
radial Schrodinger equation for the HA~\cite{Nieto,GerryKiefer}.
These states are expressed in  elementary
functions of $r$, are $SO(2, 1)$-invariant, and minimize the uncertainty
relation for the suitably defined operators $X$ and $P$.

Thus, the attempts to establish the quantum--classical
correspondence for the 3D HA, using the HO as a model, failed. In fact,
an extremely simple form of this correspondence for the HO is a
consequence of the fact that its energy levels are equidistant; as a
result, such simple correspondence cannot exist for the HA. This circumstance
has been recently mentioned in~\cite{Bellomo}; however, this reasoning
can be found
even in one of N. Bohr's fundamental papers on quantum mechanics~\cite{Bohr}
(with reference to Heisenberg and Darwin).

Therefore, in our opinion, the formulation of
quantum--classical correspondence for the HA should be based
on mathematical principles rather than on the dynamical considerations. Recently,
two approaches have been proposed in this direction. The first approach
suggested by Klauder~\cite{Klauder} is based on his continuous-representation
theory~\cite{KlauderI-II} and later was developed in a number of papers
(see~\cite{Crawford} and references therein). This approach implies that
the CS system should obey the following  requirements: (i) the dependence on
the parameters is continuous, (ii) the resolution of the identity takes place,
and (iii) the Hamiltonian yields the evolution in parameter space.
The disadvantages of this approach include the infiniteness of
the evolution in the parameter space  which is inconsistent with periodicity of the
corresponding classical motion, and the absence of a closed-form
expression for the CS in configuration space.

The another approach goes back to the Mostowski paper~\cite{Most}  and
is based on the Perelomov's method~\cite{Perel} of constructing the CS
for an  arbitrary Lie group; any dynamical considerations are avoided. In the
context of this approach, we constructed a CS system for the discrete
spectrum of HA~\cite{lett}; in particular, this system corresponds to the coset space
$SO(3, 2)/(SO(3)\otimes SO(2))$ and is expressed in the closed form in the
configuration space.

In this paper, we present the construction of this CS system  in more detail
and show that the difficulties associated with minimization of the
uncertainty relation for the HA CS noted in~\cite{Zlatev}
 can be overcome if one uses correct definitions of both.
We also show that the CS system constructed
in~\cite{lett} admits a natural generalisation to the continuous spectrum of HA;
 in this case the symmetry group is $SO(4, 1)$. Previously
a CS system for the $Sp(1,{\Bbb R})/U(1)$ space and for a continuous
series of the $Sp(1,{\Bbb R})$
group representations was constructed in a different way~\cite{Hongoh}.
Discussion of some other properties of CS for the discrete spectrum of
HA and their comparison with the HO CS properties may be found
in~\cite{lett}; it is noteworthy that notations used here coincide only partly
with those in~\cite{lett}. Concerning the symmetric space which corresponds to the CS system for
the HA discrete spectrum, see~\cite{phase}.

The plan of the present paper is the following. In section 2.1 we
consider the Kustaanheimo--Stiefel transformation and its relation to
wave functions of the discrete spectrum of HA. Since the
Kustaanheimo--Stiefel transformation and its generalisations have been
considered in many papers (see e.g.~\cite{Gerry,Nouri,Komarov,KRN}  and references
therein), we present here the needed  information only and mostly  will
follow the approach outlined in~\cite{Gerry,Cornish,BRasm}. In section 2.2
we give
the group-theoretical treatment of the HA discrete spectrum,  mostly
following~\cite{BRasm,BSW}. In section 3.1 we consider the
$SO(3, 2)$ group and the corresponding $SO(3, 2)/(SO(3)\otimes   SO(2))$
 symmetric space. In section 3.2 a CS system for the discrete spectrum is constructed; CS for the 1D
HA constructed in~\cite{GerryKiefer} are the particular case of our CS.
In section 3.3  we show that this system is a $SO(3, 2)$-invariant system of  the Perelomov's CS.
In section 2.4 we show that this CS system minimizes the so-called Robertson
inequality for 4D coordinates and momenta; this inequality is a
generalisation of the Heisenberg uncertainty relation to the case of $n$
variables. In this respect the constructed CS system is similar to the
usual CS for the Heisenberg--Weyl group. In section 4 the
continuous spectrum is considered. In section 4.1 we consider the
$SO(4, 1)$ group and its action on ${\Bbb R}^3$;
concerning the conformal action of orthogonal groups over euclidean
and pseudoeuclidean spaces see also~\cite{Thomova}. In section 4.2 the  wave
functions of the continuous spectrum of HA are considered. In section
4.3 we follow the ideas of~\cite{Atak} concerning  the Mellin
transform of the confluent hypergeometric functions to construct the CS system
for the continuous spectrum of HA; this system is similar to that
constructed in section 3.2. In section 5 it is shown that CS systems
for the discrete and continuous spectra of HA are  particular cases
of a function which interwines the different irreducible representations of the
$SO(4, 2)$ group (for more details on this group and its
representations see~\cite{Barut95,Coq} and references therein). We also
establish a relation between the results obtained above and that reported
recently~\cite{Kay} concerning  representation of the HA wave functions
in terms of the classical motion.

\section{Preliminaries}

\subsection{The Kustaanheimo-Stiefel transformation}

Let us introduce the four-vector
\[ n_{\bi x}^{\mu }=(r,{\bi x})\qquad n_{\bi x}\cdot n_{\bi x}=0\qquad n_{\bi
x}^{0}\geq 0 \qquad \mu,\nu,\ldots=0,\ldots,3.  \]
Denote the Cartesian coordinates in ${\Bbb R}^4$ as
$\xi_\alpha ,\eta_\alpha,\ \alpha,\beta=1,2$.
We also need two other coordinate systems in
 ${\Bbb R}^4$: $\xi,\eta,\varphi_\xi, \varphi_\eta$ and the
complex two-dimensional coordinates defined by
\begin{equation}\label{z1z2}
\eqalign{
z_1 =\frac{1}{\sqrt{2}}(\xi_1 +\i \xi_2) =\frac{1}{\sqrt{2}}\xi
\e^{i\varphi_\xi} \\
z_2 =\frac{1}{\sqrt{2}}(\eta_1 +\i \eta_2) =\frac{1}{\sqrt{2}}\eta
\e^{i\varphi_\eta}.
}
\end{equation}
Then the Kustaanheimo-Stiefel transformation takes the form
\[ n_{\bi x}^\mu =r_0 \bar{z}_\alpha (\sigma^\mu)_{\alpha\beta} z_\beta
\qquad \sigma^\mu =(1,\bsigma). \]
Then the Schro\"odinger equation for the 3D HA
\begin{equation}\label{1}
\left(-\frac{\hbar^2}{2\mu}\frac{\partial^2}{\partial {\bi x}^2}-
\frac{e^2}{r}\right)\psi=E\psi
\end{equation}
may be rewritten in the four-dimensional form as
\begin{eqnarray}\label{2}
\left[
\left( -\frac{\hbar^2}{2m}
\frac{\partial^2}{\partial\xi_\alpha \partial\xi_\alpha}
+\frac{1}{2}m\omega^2 \xi_\alpha \xi_\alpha
\right)
+(\xi\rightarrow\eta)
\right]\psi=\varepsilon\psi \\
\left( \frac{\partial}{\partial\varphi_\xi}+
\frac{\partial}{\partial\varphi_\eta} \right)\psi=0 \label{3}
\end{eqnarray}
where the following notations are introduced:
\[ m=4\mu \qquad \varepsilon=\frac{e^2}{2r_0}
\qquad r_0 =\hbar^2/(\mu e^2) \qquad \omega=(-E/2\mu)^{1/2}.\]
Let a solution
$\psi=\psi (\xi,\eta,\varphi_\xi,\varphi_\eta)$
of the equations~(\ref{2}),(\ref{3}) is known, then we can obtain
the solution of equation~(\ref{1}) setting
$\varphi_\xi =-\varphi_\eta=\varphi/2$.
Then the Kustaanheimo--Stiefel transformation reduces  to  usual
transformation to the parabolic coordinates
\begin{eqnarray}
x_1 =r_0\xi\eta \cos \varphi \qquad x_2 =r_0\xi\eta \sin\varphi \nonumber \\
x_3 =\frac{r_0}{2}(\xi^2 -\eta^2 ) \qquad
r =\frac{r_0}{2}(\xi^2 +\eta^2 ) .\nonumber
\end{eqnarray}
Let $E<0$; then rescaling the coordinates we can reduce equations~(\ref{2}),(\ref{3})
 to two Schrodinger equations for two 2D HO with unit mass
and frequency and with the same values of angular momentum. Then
the  functions
\begin{eqnarray}
\Psi_{n_1 n_2 m}({\bi x})=\psi_{n_1 n_2 m}({\bi x}/(nr_0))\nonumber \\
n=n_1 +n_2 +|m|+1 \qquad E=-\frac{\varepsilon}{n^2} \nonumber
\end{eqnarray}
obey equation~(\ref{1}), where
\[ \psi_{n_1 n_2 m}({\bi x})\equiv |n_1 n_2 m\rangle=
(-1)^{n_{1}+\frac{1}{2}(m-|m|)}\frac{\e^{\i m\phi }}{\pi^{1/2}}
\varphi_{n_1 |m|}(\xi)\varphi_{n_2 |m|}(\eta) \]
and
\[ \varphi_{n|m|}(\xi)=\e^{-\xi^2 /2}\xi^{|m|}
\left(\frac{n!}{(n+|m|)!} \right)^{1/2} L_n^{|m|}(\xi^2)\]
is a solution of the radial Schrodinger equation for the 2D HO with unit
mass and frequency and with the angular momentum equal to $|m|$. The
above solutions are real and normalized as
\[\int_{0}^\infty \d (\xi^2)\,
\varphi_{n|m|}(\xi) \varphi_{n'|m|}(\xi)=\delta_{nn'}.\]
Then taking into account that
$r^{-1}\d x_1 \d x_2 \d x_3 =\d (\xi^2) \d (\eta^2) \d\varphi$ we obtain
\begin{equation}\label{product}
\fl \langle n_1 n_2 m|{n'}_1 {n'}_2 m'\rangle \equiv
\int_{{\Bbb R}^3} r^{-1}\d x_1 \d x_2 \d x_3 \, \psi_{n_1 n_2 m}({\bi x})
\psi_{{n'}_1 {n'}_2 m'}({\bi x})=\delta_{n_1 {n'}_1} \delta_{n_2 {n'}_2}
\delta_{mm'}.
\end{equation}
The measure in the right-hand side of the above expression is just a
Lorentz-invariant measure over the light cone  $n_{\bi x} \cdot n_{\bi x}=0$.

\subsection{The dynamical symmetry group}
Let us introduce the operators $a_\alpha , b_\alpha$  and their Hermitean conjugates as
 \begin{equation}\label{ab-z}
\eqalign{
a_\alpha=\frac{1}{\sqrt{2}}({\bar{z}}_\alpha +\partial_{z_\alpha}) \qquad
a_\alpha^\dagger =\frac{1}{\sqrt{2}}(z_\alpha -\partial_{{\bar{z}}_\alpha})
\\
b_\alpha=\frac{1}{\sqrt{2}}\varepsilon_{\alpha\beta}
(z_\beta +\partial_{{\bar{z}}_\beta}) \qquad
b_\alpha^\dagger =\frac{1}{\sqrt{2}}
\varepsilon_{\alpha\beta}({\bar{z}}_\beta -\partial_{z_\beta}).
}
\end{equation}
Then nonvanishing commutation relations are
\[ [a_\alpha, a^\dagger_\beta ]= [b_\alpha, b^\dagger_\beta
]=\delta_{\alpha\beta}.\]
Then we can express the vectors $|n_1 n_2 m\rangle$ as
\begin{equation}\label{6}
\eqalign{
|n_1 n_2 m\rangle= [n_1 ! (n_1 +|m|)! n_2 ! (n_2+|m|)!]^{-1/2} \\
\times\left\{
\begin{array}{ll}
a_1^{\dagger n_2+|m|} a_2^{\dagger n_1} b_1^{\dagger n_1+|m|}
b_2^{\dagger n_2} |0\rangle & m\geq 0 \\
a_1^{\dagger n_2} a_2^{\dagger n_1+|m|} b_1^{\dagger n_1}
b_2^{\dagger n_2 +|m|} |0\rangle & m\leq 0
\end{array}
\right.
}
\end{equation}
where
\begin{equation}\label{6a}
|0\rangle =\exp (-z_\alpha \bar{z}_\alpha) \qquad
a_\alpha |0\rangle =b_\alpha |0\rangle =0 \qquad \alpha=1,2.
\end{equation}
Then the linear shell of the vectors $|n_1 n_2 m\rangle$
 may be considered as a
subspace in the Fock space of a bosonic system of four degrees of
freedom; this subspace is defined by the constraint
\begin{equation}\label{Hphys}
(a_\alpha a^\dagger_\alpha -b_\alpha b^\dagger_\alpha)|{\rm phys}\rangle=0
\end{equation}
which, as can be readily seen, coincides with~(\ref{3}). We denote this
subspace as  $H_{\rm phys}.$

From~(\ref{6}) it follows that
the following representation of the algebra $so(4,2)$
\begin{equation}\label{8}
\eqalign{
L_{ij}=\frac{1}{2}(a^\dagger \sigma_k a +b^\dagger \sigma_k b)
\qquad L_{i5}=-\frac{1}{2}(a^\dagger\sigma_i C b^\dagger -a C\sigma_i b) \\
L_{i0}=\frac{1}{2\i}(a^\dagger\sigma_i C b^\dagger + a C\sigma_i b) \qquad
L_{50}=\frac{1}{2}(a^\dagger a+b^\dagger b+2)   \\
L_{i6}=-\frac{1}{2}(a^\dagger \sigma_i a -b^\dagger \sigma_i b) \qquad
L_{56}=(\i/2)(a^\dagger Cb^\dagger -aCb) \\
L_{60}=\frac{1}{2}(a^\dagger Cb^\dagger +aCb)
}
\end{equation}
may be defined in $H_{\rm phys}$, where
 $C=\i \sigma _{2}$ and the generators $L_{AB},\ A,B,\ldots=0,\ldots,3,5,6$
obey commutation relations
\begin{equation}
[L_{AB},L_{CD}]=\i (\eta _{AD}L_{BC}+\eta _{BC}L_{AD}-\eta
_{AC}L_{BD}-\eta _{BD}L_{AC})  \label{9}
\end{equation}
where $\eta _{AB}={\rm diag}(+1,-1,-1,-1,+1,-1)$. Then from~(\ref{6})
and~(\ref{8}) it follows that
\begin{equation}\label{10}
L_{50}|n_{1}n_{2}m\rangle =(n_{1}+n_{2}+|m|+1)|n_{1}n_{2}m\rangle .
\end{equation}
Substituting~(\ref{ab-z}) into~(\ref{8})
we obtain the following expressions for
generators $L_{AB}$  in the configuration space:
\begin{equation}\label{11}
\eqalign{
L_{ij}=\varepsilon_{ijk} ({\bi x}\times {\bi p})_k \qquad
L_{i6}=-\frac{1}{2}x_i p^2 +p_i ({\bi x}{\bi p})+\frac{1}{2}x_i \\
L_{i5}=-\frac{1}{2}x_i p^2 +p_i ({\bi x}{\bi p})-\frac{1}{2}x_i   \qquad
L_{65}=({\bi x}{\bi p})-\i \\
L_{i0}=-rp_i \qquad L_{60}=\frac{1}{2}(rp^2 -r) \qquad
L_{50}=\frac{1}{2}(rp^2 +r).
}
\end{equation}
The generators $L_{\mu\nu}$ induce the Lorentz transformations of four-vectors
 $n_{\bi x}^\mu$.

\section{Discrete spectrum}

\subsection{The $Sp(2,{\bf R})/U(2)$ space}

Let us consider a set of complex symmetric $2\times 2$ matrices  obeying the condition
\begin{equation}\label{12a}
I-\Lambda\Lambda^\dagger >0.
\end{equation}
On these matrices we can define the action of the $Sp(2,{\Bbb
R})$ group so that they become the symmetric space
\[ SO(3,2)/(SO(3)\otimes SO(2))\simeq Sp(2,{\Bbb R})/U(2).\]
This space has been considered in detail previously (see e.g.~\cite{Balbinot}).
Introduce a three-vector $\bi u$ as  $\Lambda=C\bsigma {\bi u}$.
Then~(\ref{12a}) is equivalent to the conditions
\begin{equation}\label{12}
|{\bi u}^2|<1 \qquad 1-2{\bi u}{\bi u}^* +{\bi u}^2 {\bi u}^{*2}>0.
\end{equation}
The infinitesimal operators corresponding to the action of the $SO(3, 2)$
group over this space are given by~\cite{Balbinot}
\begin{equation}\label{13}
\eqalign{
L_{ij}=\i \left(u_i\frac{\partial}{\partial u_j}-
u_j \frac{\partial}{\partial u_i} \right) \qquad
L_{50}={\bi u}\frac{\partial}{\partial {\bi u}} \\
L_{5i}=\i \left( \frac{1+{\bi u}^2}{2}\frac{\partial}{\partial u_i}-
u_i \left({\bi u}\frac{\partial}{\partial {\bi u}} \right) \right) \\
L_{0i}=- \left( \frac{1-{\bi u}^2}{2}\frac{\partial}{\partial u_i}+
u_i \left({\bi u}\frac{\partial}{\partial {\bi u}} \right) \right) .
}
\end{equation}
Let us introduce a unit complex four-vector as
\begin{equation}\label{14}
k_{\bi u}^{\mu }=\left( \frac{1+{\bi u}^{2}}{1-{\bi u}^{2}},
\frac{2{\bi u}}{1-{\bi u}^{2}}\right) \qquad k_{\bi u} \cdot k_{\bi u}=1.
\end{equation}
Then we can rewrite the conditions~(\ref{12}) as
\begin{equation}
w_{\bi u}^0 >0 \qquad
w_{\bi u}\cdot w_{\bi u}=\frac{1-2{\bi u}{\bi u}^{*}+{\bi u}^{2}{\bi u}^{*2}
}{|1+{\bi u}^{2}|^{2}}>0   \label{15}
\end{equation}
where  $w_{\bi u}^{\mu }=\mathop{{\rm Re}}k_{\bi u}^{\mu}$. The action
of generators $L_{\mu\nu}$~(\ref{13}) corresponds to the Lorentz
transformations of the vector $k^\mu_{\bi u}$.

\subsection{Coherent states}

Let $\bi u$ be a complex three-vector having the components
\begin{equation}
{\bi u}=\left( \frac{\i }{2}(\lambda _{2}-\lambda _{1}),\frac{1}{2}
(\lambda _{1}+\lambda _{2}),0\right)   \label{16}
\end{equation}
and satisfying the conditions~(\ref{12}). We now construct the
superposition of states
\begin{equation}
|{\bi u}\rangle =c_{0}\sum\limits_{n=0}^{\infty
}\sum\limits_{m=-\infty }^{\infty }(\lambda _{1}\lambda _{2})^{\frac{1}{2}
(2n+|m|+1)}\left( \frac{\lambda _{1}}{\lambda _{2}}\right) ^{m/2}|nnm\rangle.
\label{17}
\end{equation}
Using the formulas~\cite{HTF-2}
\begin{eqnarray}
\sum\limits_{n=0}^{\infty }\frac{n!}{\Gamma (n+\alpha +1)}L_{n}^{\alpha
}(x)L_{n}^{\alpha }(y)z^{n}  \nonumber \\
=(1-z)^{-1}\exp \left( -z\frac{x+y}{1-z}\right) (-xyz)^{-\alpha
/2}J_{\alpha }\left( 2\frac{(-xyz)^{1/2}}{1-z}\right) \qquad |z|<1  \nonumber
\\
\bs\sum\limits_{n=-\infty }^{\infty }t^{n}J_{n}(z) =\exp \left[
(t-t^{-1})z/2\right]  \label{17a}
\end{eqnarray}
we obtain
\begin{equation}\label{18}
\langle {\bi x}|{\bi u}\rangle =\frac{c_{0}}{2\sqrt{\pi }}
({\bi k}_{\bi u}^{2})^{1/2}\exp (- k_{\bi u}\cdot n_{\bi x}).
\end{equation}
It can be readily seen that an arbitrary three-vector satisfying
conditions~(\ref{12}) can be obtained by applying the $SO(3)$ transformations to a
certain three-vector defined by~(\ref{16}). Due to~(\ref{11})
such a transformation corresponds to certain transformation in $H_{\rm phys}$.
 Then the vector $|{\bi u}\rangle$ defined
by the right-hand side of equality~(\ref{18}) can be represented as a
superposition of vectors of the space $H_{\rm phys}$
   for an arbitrary $\bi u$ which obeys the conditions~(\ref{12}). Then hereafter we will consider
$\bi u$ as an arbitrary element of the space $Sp(2,{\Bbb R})/U(2)$.

Let us choose the normalization constant so that
$\langle {\bi u}|{\bi u}\rangle=1$ i.e.
\[ |c_{0}|^{2}=\frac{1-2{\bi u}{\bi u}^{*}+{\bi u}^{2}{\bi u}^{*2}}
{|{\bi u}^{2}|}.  \]
Thus, both the conditions~(\ref{12}) are necessary; the first one is necessary for
convergence of the series~(\ref{17}) and the second one for normalizability of the
resulting expression. Then using~(\ref{15}) we finally obtain
\begin{equation}\label{20}
\langle {\bi x}|{\bi u}\rangle =\frac{1}{\pi ^{1/2}}(w_{\bi u}\cdot w_{\bi
u})^{1/2}\exp (- k_{\bi u}\cdot n_{\bi x}).
\end{equation}
CS for the 1D HA constructed in~\cite{GerryKiefer} may be easily obtained
as a particular case of~(\ref{20}) putting $x^1 =x^2 =k^1_{\bi u}=
k^2_{\bi u} =0$.

\subsection{Symmetry properties}

Now we show that the system  $|{\bi u}\rangle$ is $SO(3, 2)$-invariant. This
system is obviously $SO(3, 1)$-invariant; on the other hand, we have the equality
\[ \e^{\i \varepsilon L_{50}}|{\bi u}\rangle =
\e^{\i\varepsilon}|{\bi u}e^{\i\varepsilon }\rangle  \]
 which can be proven either using~(\ref{10})  and~(\ref{17}) or in
the infinitesimal form using~(\ref{11}) and~(\ref{13}).
Then the full $SO(3, 2)$-invariance of the system  follows then from the commutation relations~(\ref{9}).

From here it follows that the system $|{\bi u}\rangle$ is a system of  Perelomov's CS for
the $SO(3, 2)$ group constructed starting from the $SO(3)\times
SO(2)$-invariant vector $|0\rangle$.
This fact also can be  proven directly for the following particular cases:
\begin{equation}\label{22}
\eqalign{
1. \ \Lambda=\left(
\begin{array}{ll}
0 & \alpha \\
\alpha & 0
\end{array}
\right)   \\
2.\ \Lambda=\left(
\begin{array}{ll}
\alpha_1 & 0\\
0 & \alpha_2
\end{array}
\right) \qquad \alpha_1 \alpha_2=0.
}
\end{equation}
To this end let us introduce the operators $A_\alpha ,B_\alpha$ as
\begin{equation}\label{AB}
 a_\alpha =A_\alpha +\i B_\alpha \qquad b_\alpha =A_\alpha -\i B_\alpha.
\end{equation}
Since the matrices $C\sigma_i$
are symmetric, then from~(\ref{8}) it follows that
generators of the $SO(3, 2)$ subgroup of the $SO(4, 2)$ group
may be represented as a nondegenerate linear combination of generators
of the $Sp(2,{\Bbb R})\simeq SO(3,2)$ group:
\begin{equation}\label{23}
\eqalign{
X_{\alpha \beta }=A_{\alpha }A_{\beta }+ B_{\alpha }B_{\beta }\qquad
X_{\alpha \beta }^{\dagger}=
A_{\alpha }^{\dagger }A_{\beta }^{\dagger }+
B_{\alpha }^{\dagger }B_{\beta }^{\dagger } \\
Y_{\alpha \beta }=
\frac{1}{2}(A_{\alpha}A_{\beta}^{\dagger}+
A_{\beta}^{\dagger }A_{\alpha })+
\frac{1}{2}(B_{\alpha }B_{\beta }^{\dagger }+
B_{\beta }^{\dagger}B_{\alpha}).
}
\end{equation}
It follows from~(\ref{23}) that the group $Sp(2,{\Bbb R})$ is a group of canonical
transformations of  operators  $A_\alpha ,A^\dagger_\alpha$ and $B_\alpha ,B^\dagger_\alpha$
separately. Then since~(\ref{6a}) is equivalent to
 \[ A_\alpha |0\rangle =B_\alpha |0\rangle =0\]
then we can use the analogy with the usual CS for a bosonic system
of two degrees of freedom~\cite{Perel} to obtain the equalities
\begin{equation}\label{25}
(A_\alpha -\Lambda_{\alpha\beta}A^\dagger_\beta)|\Lambda\rangle =
(B_\alpha -\Lambda_{\alpha\beta}B^\dagger_\beta)|\Lambda\rangle =0.
\end{equation}
Denote as ${\cal H}_A$ the Hilbert space of states of bosonic system of
two degrees of freedom composed by  vectors of the form
$A^\dagger_{\alpha_1}\ldots A^\dagger_{\alpha_n}|0\rangle_A$, where
$A_\alpha |0\rangle_A=0, \ \alpha=1,2$. Following~\cite{Perel}, define
in ${\cal H}_A$ the CS system for the space $Sp(2,{\Bbb R})/U(2)$ as
\[ |\Lambda\rangle_A =[\det (I-\Lambda\Lambda^\dagger )]^{1/4}
\exp\left(\frac{1}{2} \Lambda_{\alpha\beta}
A^\dagger_\alpha A^\dagger_\beta\right)|0\rangle_A . \]
The space ${\cal H}_B$ and its CS $|\Lambda\rangle_B$ may be defined in
the completely analogous way. Then we can consider the
representation~(\ref{23}) of the $SO(3,2)$ group as acting in the subspace
of the space ${\cal H}_A \times {\cal H}_B$ defined by the
constraint~(\ref{Hphys}). Consider in the space
${\cal H}_A \times {\cal H}_B$ the system of states
\begin{equation}\label{AxB}
|\Lambda\rangle =|\Lambda\rangle_A \otimes |\Lambda\rangle_B =
[\det (I-\Lambda\Lambda^\dagger )]^{1/2}
\exp\left(\frac{1}{2} \Lambda_{\alpha\beta}X^\dagger_{\alpha\beta}\right)
|0\rangle
\end{equation}
where $X_{\alpha\beta}$ are given by~(\ref{23}). Using~(\ref{25}) it is
easy to show that the vectors $|\Lambda\rangle$ obey the
constraint~(\ref{Hphys}) and then belong to ${\cal H}_{\rm phys}$. From
here it follows that~(\ref{AxB}) is true Perelomov HA CS system for the
space $Sp(2,{\Bbb R})/U(2)$ and then must coincite with~(\ref{20}).
Using the equality~(\ref{6}) above and the formula~\cite{HTF-2}
\[ \sum_{n=0}^\infty L_n^\alpha (x) z^n=(1-z)^{-\alpha-1}\exp
\left(\frac{xz}{z-1}\right) \qquad |z|<1 \]
we can directly prove that~(\ref{AxB}) indeed coincides with~(\ref{20})
to within a phase multiplier if $\Lambda$ has the form~(\ref{22}).

\subsection{Robertson relations}

Introduce the Hermitean operators $Q_a ,\ a=1,\ldots,8$ as
\[\eqalign{
Q_\alpha =\xi_\alpha \qquad
Q_{\alpha+2}=-\i\frac{\partial}{\partial \xi_\alpha}  \\
Q_{\alpha+4} =\eta_\alpha \qquad
Q_{\alpha+6}=-\i\frac{\partial}{\partial \eta_\alpha}
}\]
and define their dispersion in a given state as
\[ \Sigma_{ab}=\frac{1}{2}\langle Q_a Q_b +Q_b Q_a \rangle -\langle Q_a
\rangle \langle Q_b \rangle .\]
By the virtue of~(\ref{z1z2}),(\ref{ab-z}), (\ref{AB}) and (\ref{25}) the operators $Q_a$ acting on the
vectors $|{\bi u}\rangle$ satisfy $4 = 8 / 2$  linearly independent equalities. The
Robertson inequality for the dispersion of a set of Hermitean operators
\[ \det \Sigma \geq \det\Omega \qquad \Omega_{ab}=-\frac{\i}{2}
\langle [Q_a, Q_b]\rangle \]
in view of the results of~\cite{Trifonov}
is then transformed into the equality,  if the mean values are taken in an arbitrary CS $|{\bi u}\rangle$.

\section{Continuous spectrum}

\subsection{The $SO(4,1)$ group}

Let us introduce the generators
\[ \bPi^\pm_i =L_{6i}\pm L_{0i}. \]
The generators $\bPi^+$ and $\bPi^-$
form two Abelian subgroups, which we denote by
${\cal T}^+$ and ${\cal T}^-$; the subgroups induced by the generators
$L_{06}$ and $L_{ij}$ we denote as ${\cal T}^0$ and $\cal R$
respectively. Finite transformations are denoted as
\[
\Theta_{\pm}({\bi a})= \exp (\i\bPi^{\pm}{\bi a}) \qquad
\Theta_{0}(\varepsilon )= \exp (\i L_{06} \varepsilon ).\]
Consider the action ${\bi v}\mapsto {\bi v}_g$
of elements $g\in SO(4,1)$ over the vectors
 ${\bi v}\in {\Bbb R}^3$ defined by
\begin{eqnarray}
 g=\Theta_{-}({{\bi a}}):\ {{\bi v}}_{g}={{\bi v}}-{{\bi a}}\nonumber \\
\label{cont1}
g=\Theta_{+}({{\bi a}}):\ {{\bi v}}_{g}=
\frac{{{\bi v}}+{{\bi a}}{v}^{2}}{1+2{\bi v}{\bi a}+{v}^{2}{a}^{2}} \\
g=\Theta_{0}(\varepsilon ):\ {{\bi v}}_{g}={{\bi v}}\e^{\varepsilon}.
\nonumber
\end{eqnarray}
Generators have the form
\begin{equation}\label{cont2}
\eqalign{
\i\bPi^{-}=-\frac{\partial}{\partial {{\bi v}}} \qquad
\i\bPi^{+}={v}^{2}\frac{\partial}{\partial {{\bi v}}}-2{{\bi v}}
\left( {{\bi v}}\frac{\partial}{\partial {{\bi v}}}\right) \\
\i L_{06}= {{\bi v}}\frac{\partial}{\partial {{\bi v}}} \qquad
\i L_{ik}={v}_{k}\frac{\partial}{\partial  {v}_{i}}-
{v}_{i}\frac{\partial}{\partial {v}_{k}}.
}
\end{equation}
The stationary subgroup of the point
${\bi v}={\bi 0}$ is ${\cal K}={\cal T}^+
\circledS ({\cal T}^0 \otimes {\cal R})$; then the space ${\Bbb R}^3$
equipped with such an action of the $SO(4, 1)$ group
may be identified with the coset space $SO(4,1)/{\cal K}$.
 Then the action of the group $SO(4, 1)$ over the unit real four-vector
$k_{\bi v}$~(\ref{14}) is defined; the generators $L_{\mu\nu}$
correspond to the Lorentz transformations of this vector.

\subsection{Wave functions}
In the case of positive energy we can use the coordinate rescaling to reduce
equations~(\ref{2}),(\ref{3}) to the Schrodinger equations for two `oscillators' having
unit mass, frequency equal to i, and the same values of angular momentum.
Then the solutions of  equation~(\ref{1}) corresponding to  energy
$E=\varepsilon (\rho_1 +\rho_2)^{-2}$ are
\[ \Psi_{\rho_1 \rho_2 m}({\bi x})=\psi_{\rho_1 \rho_2 m}
({\bi x}(r_0(\rho_1 +\rho_2))^{-1}) \]
where
\begin{eqnarray}
\psi_{\rho_1 \rho_2 m}({\bi x})\equiv |\rho_1 \rho_2 m\rangle=
\e^{\i m\phi}
\varphi_{\rho_1 |m|}(\xi)\varphi_{\rho_2 |m|}(\eta) \nonumber \\
\bs\varphi_{\rho |m|} (\xi)=(2\pi\i\e^{\pi\rho})^{-1/2}
\left|\Gamma\left(-\i\rho+\frac{|m|+1}{2}\right) \right| \nonumber \\
\times \e^{\i\xi^2 /2}(-\i\xi^2)^{|m|/2}\mathop{_1 F_1 \left(
-\i\rho+\frac{|m|+1}{2} ,|m|+1,-\i \xi^2 \right) } . \nonumber
\end{eqnarray}
To  within the change  $|m|$ to $2l + 1$  the functions $\varphi_{\rho |m|}$ coincide
with radial components of wave functions of the continuous spectrum of HA
obtained previously in~\cite{Szmytk}. The normalization
factors are chosen so that the functions $\varphi_{\rho |m|} (\xi)$
 are real and satisfy the normalization conditions
\[\int_{0}^\infty \d (\xi^2)\,
\varphi_{\rho|m|}(\xi) \varphi_{\rho'|m|}(\xi)=\delta (\rho-\rho').\]
Then
\[ \langle \rho_1 \rho_2 m|{\rho'}_1 {\rho'}_2 m'\rangle=
\delta(\rho_1 -{\rho'}_1) \delta(\rho_2 -{\rho'}_2)\delta_{mm'}.\]
The equality
\begin{equation}\label{cont6}
L_{06} |{\rho}_1 {\rho}_2 m\rangle =-(\rho_1 +\rho_2)
|{\rho}_1 {\rho}_2 m\rangle
\end{equation}
holds.

\subsection{Coherent states}

Let ${\bi v}=(-v\cos\theta ,v\sin\theta,0)$. By the analogy with~(\ref{17})
define the states $|{\bi v}\rangle$ as
\begin{equation}\label{cont3}
|{\bi v}\rangle =({\bi k}_{\bi v}^{2})^{-1/2}
 \sum\limits_{m=-\infty}^\infty \e^{\i m\theta}
\int\limits_{-\infty}^\infty \d\rho \, v^{-2\i\rho}|\rho\rho m\rangle.
\end{equation}
Inverting the Mellin transform  of the Bessel function~\cite{IT-1}
\begin{eqnarray}
\int\limits_0^\infty t^{s-1} \d t\, (1+t^2)^{-1}\exp\left(-\i
\frac{\xi^2 +\eta^2}{2}\frac{1-t^2}{1+t^2}\right) J_{|m|}
\left(-\i\xi\eta\frac{2t}{1+t^2}\right)\nonumber \\
=\pi\i^{|m|+1}\e^{\pi\rho}
\left. \varphi_{\rho |m|}(\xi)\varphi_{\rho |m|}(\eta)
\right|_{\i\rho=\frac{s-1}{2}} \nonumber
\end{eqnarray}
and using~(\ref{17a}) we obtain
 \begin{equation}\label{|v>}
\langle {\bi x}|{\bi v}\rangle =\exp (-\i k_{\bi v}\cdot n_{\bi x}).
\end{equation}
Using~(\ref{cont6}) and~(\ref{cont3}) we obtain that at $g\in {\cal T}^0$
\begin{equation}\label{cont5}
T(g)|{\bi v}\rangle =
\left(\frac{\d\mu (k_{{\bi v}_g})}{\d\mu (k_{\bi v})} \right)^{1/3}
 |{\bi v}_g \rangle
\end{equation}
where $T(g)$ is representation of the $SO(4, 1)$ group with the Lie algebra  given by~(\ref{11})
and  $\d\mu (k)=(k^0)^{-1}\d^3 {\bi k}$ is the Lorentz-invariant
measure on the hyperboloid $k\cdot k=1$. The
equality~(\ref{cont5}) can be also proved in the infinitesimal form
using~(\ref{11}) and~(\ref{cont2}). From the other hand, the validity
of~(\ref{cont5}) at
$g\in SO(3,1)$ is obvious, so it is correct for all $g\in SO(4,1)$.

Define the space $SO(4,1)/({\cal T}^+
\circledS {\cal R})$ as a set of pairs $({\bi v},\tau)$, where
$\tau\in {\Bbb R}\backslash \{ 0\}$ and the action of the $SO(4,1)$
group is defined by~(\ref{cont1}) and
\[ \tau_g =
\left(\frac{\d\mu (k_{{\bi v}_g})}{\d\mu (k_{\bi v})} \right)^{1/3} \tau.\]
Then the states
\[ |{\bi v}\tau\rangle =\tau^{-1}|{\bi v}\rangle \]
compose a system of  Perelomov's CS for the mentioned space.

\section{Relation to the conformal group}

The twistor space $SO(4,2)/(SO(4)\otimes SO(2))$ is a domain in
${\Bbb C}^4$ defined by the inequalities
\begin{equation}\label{tw1}
|u_a u_a |<1 \qquad 1-2u_a u_a^* +|u_a u_a|^2 >0 \qquad a=1,\ldots,4.
\end{equation}
We obtain another realization of the twistor space considering the
mapping
\begin{equation}\label{tw2}
z^0 =\i\frac{1+u_a u_a -2u_4}{1-u_a u_a +2\i u_4} \qquad
z^k =\frac{2\i u_k}{1-u_a u_a +2\i u_4}.
\end{equation}
Then~(\ref{tw1}) transforms to
\[ \Im z^0 >0 \qquad \Im z \cdot \Im z >0.\]
Consider the set of all holomorphic $C^\infty$-functions which are square integrable over the twistor
sppace with respect to the measure $\d^4 z \d^4 \bar{z}$.
Over this set we can define the $SO(4, 2)$ group irreducible representation belonging to the discrete
series and having the generators
\begin{eqnarray}
\i (L_{5\mu}+L_{6\mu})
= (z\cdot z)\frac{\partial}{\partial z^\mu}-
2z_{\mu} \left(z\cdot \frac{\partial}{\partial z}\right)-2z_\mu
\nonumber \\
\i (L_{5\mu}-L_{6\mu}) =\frac{\partial}{\partial z^\mu} \label{tw3} \\
\i L_{\mu\nu}=z_\mu \frac{\partial}{\partial z^\nu} -
z_\nu \frac{\partial}{\partial z^\mu} \qquad
\i L_{65}=z\cdot \frac{\partial}{\partial z}+1. \nonumber
\end{eqnarray}
Consider the functions
\[ \langle {\bi x}|z\rangle = \e^{\i n_{\bi x}\cdot z}.\]
Then one can show~\cite{JV} that the integral transform
\begin{equation}\label{tw4}
F(z)=\int \frac{\d^3 {\bi x}}{r}\langle {\bi x}|z\rangle f({\bi x})
\end{equation}
interwines the representations ~(\ref{11}) and~(\ref{tw3}) of the
$SO(4, 2)$ group. At the
level of Lie algebras this fact can be directly observed since
 the difference between generators~(\ref{11})
and~(\ref{tw3}) vanishes acting on the functions $\langle {\bi x}|z\rangle$.

We can pass from the twistor space to the
$SO(3,2)/SO(3)\times SO(2)$ space letting $u_4 =0$; then~(\ref{tw1}) transforms  into~(\ref{12}),
and from~(\ref{tw2}) it follows that $z^\mu =\i k^\mu_{\bi u}$. From the other hand, the functions
$\langle {\bi x}|z\rangle$ transform into CS given by~(\ref{20}) to within a normalization factor.

Letting $\Im z^\mu =0$ we pass to the Shilov boundary of the twistor space which
coincides with the Minkowski space.
If we additionally let $z^\mu =k^\mu_{\bi v}$ then the functions
$\langle {\bi x}|z\rangle$  pass into the states~(\ref{|v>}).

Passing to the Shilov boundary, the representation~(\ref{tw3})
transforms into the representation which describes massless
spin zero particles over the Minkowski space~\cite{Ruhl}. In this case, the
transform~(\ref{tw4}) shows the coincidence of representations
of the $SO(4, 2)$ group which describe the hydrogen atom and massless
spin zero particles over the Minkowski space. This coincidence has
been previously proven in a more complicated way in~\cite{MackTod}.

Let us consider now the manifold which belongs to the boundary of
the twistor space and is defined by the equality
 $\Im z\cdot \Im z =0$ (however, we still have $\Im z^0 >0$),
and moreover we assume that $z_i z_i =0$.
Then   we can represent the HA wave functions in the form of an integral
over this manifold of the functions $\langle {\bi x}|z\rangle$
with a certain weight factor~\cite{Kay}.
 This indicates the possibility of a quasi-classical description
of HA in terms of CS constructed above; this question requires
a further investigation.

\ack

I am grateful to Yu P Stepanovsky for his constant support
and helpful discussions.
The fulfillement of this work
 was made possible by
the fact that the Institute of Physics (Bristol) granted the free access
to its Electronic Journals from October 22 to December 22 2000.

\end{document}